\documentclass[nofootinbib, showkeys,iop,twocolumn]{revtex4-1}
\usepackage{amsmath}
\usepackage{graphicx}
\usepackage{placeins}

\begin{document}
% Use the \preprint command to place your local institutional report
% number in the upper righthand corner of the title page in preprint mode.
% Multiple \preprint commands are allowed.
% Use the 'preprintnumbers' class option to override journal defaults
% to display numbers if necessary
%\preprint{}

%Title of paper
\title{Limitations of rotating-wave approximation in magnetic resonance: Characterization and elimination of the Bloch--Siegert shift in magneto--optics}

% repeat the \author .. \affiliation  etc. as needed
% \email, \thanks, \homepage, \altaffiliation all apply to the current
% author. Explanatory text should go in the []'s, actual e-mail
% address or url should go in the {}'s for \email and \homepage.
% Please use the appropriate macro foreach each type of information

% \affiliation command applies to all authors since the last
% \affiliation command. The \affiliation command should follow the
% other information
% \affiliation can be followed by \email, \homepage, \thanks as well.
\author{J. I. Sudyka}
\email[]{julia.sudyka@doctoral.uj.edu.pl}
\author{S. Pustelny}
\author{W. Gawlik}

%\homepage[]{Your web page}
%\thanks{}
%\altaffiliation{}
\affiliation{Marian Smoluchowski Institute of Physics, Jagiellonian  University,  \L{}ojasiewicza 11, 30-348, Krak\'o{}w, Poland}
\keywords{Bloch--Siegert effect, magneto--optical effects, magnetic resonance, rotating wave approximation}
%\email{julia.sudyka@doctoral.uj.edu.pl} 
%Collaboration name if desired (requires use of superscriptaddress
%option in \documentclass). \noaffiliation is required (may also be
%used with the \author command).
%\collaboration can be followed by \email, \homepage, \thanks as well.
%\collaboration{}
%\noaffiliation

\date{\today}

\begin{abstract}
We present investigations of radio-frequency resonances observed in a rubidium vapor. With measurements of a systematic shift of radio-frequency resonances, induced by an off-resonance component of an oscillating field (the Bloch--Siegert shift), we demonstrate limitations of the rotating-wave approximation. Experimental results are supported with numerical calculations, reproducing all features of the observed signals. We also present the system where the effect can be controllably turned on or off (by switching from oscillating to rotating magnetic field). Besides its fundamental aspect of liberating the measurements from systematic errors in spectroscopy, magnetometry, searches for anomalous spin intaractions, etc.. This has important implications for quantum metrology, quantum information processing and new material physics.
\end{abstract}
% insert suggested PACS numbers in braces on next line
\pacs{}
% insert suggested keywords - APS authors don't need to do this
%\keywords{}

%\maketitle must follow title, authors, abstract, \pacs, and \keywords
\maketitle

% body of paper here - Use proper section commands
% References should be done using the \cite, \ref, and \label commands

\section{Introduction}
A standard treatment of a two-level system interacting with a resonant electromagnetic field is based on decomposition of an oscillating field into two counter-rotating components and consideration of system's interaction with only one of them. This approach, known as rotating-wave approximation (RWA), is justified in most cases, such as not-too-strong and optical fields. In a particular case of oscillating magnetic field $\boldsymbol{B_{RF}}\cos(\omega t)$ applied to atoms immersed in perpendicularly oriented static magnetic field $\boldsymbol{B_0}$, the interaction leads to appearance of the resonance at frequency $\Omega_r$. In RWA, $\Omega_r$ equals the static-field Larmor frequency $\Omega_0 = g \mu_B B_{0}/ \hbar$, where $g, \mu_B, \hbar$ are the gyromagnetic ratio, the Bohr magneton and the reduced Planck constant, respectively. However, due to interaction with counter-rotating component of the oscillating magnetic field, the actual resonance position is shifted from $\Omega_0$. The lowest order approximation of this shift was primarily derived by Bloch and Siegert \cite{bloch-siegert}, and is given by
 \begin{equation}
 \Delta \Omega =\Omega_0 - \Omega_r  \approx \frac{\Omega^2_{RF}}{16 \Omega_0}, 
\label{bs} 
 \end{equation}
 where $\Omega_{RF} = g \mu_B B_{RF}/ \hbar$. $\Delta \Omega$ is known as the Bloch--Siegert shift (BSS).

Since its first description in 1940 \cite{bloch-siegert}, BSS was carefully investigated theoretically \cite{cohen,shirley,hannaford} to develop analytical solution for $\Delta \Omega$. Experimental studies of the effect were performed with different electron-spin systems \cite{amoruzzi,arimondo} as well as in nuclear magnetic resonance (NMR) \cite{nmr1, nmr2}. Recently, the effect was revisited for novel applications in quantum metrology \cite{mapping} and quantum information processing \cite{qbit, qbit2}. For strong driving fields, when the shift becomes comparable with the eigenfrequency of the system, significant departures from RWA were observed in nitrogen-vacancy centers in diamonds \cite{nv1,nv2}, which may find applications for generating high-fidelity quantum gates. The effect was also exploited to control valleytronic properties of novel Dirac materials such as monolayer WS$_2$ (see Ref. \cite{valley} and references therein). Despite its growing importance, to our knowledge, no direct comparison of resonance signals induced by rotating and non-rotating fields has been reported. 

A particular example where BSS is not taken into consideration are nonlinear magneto--optical effects in alkali metal vapors \cite{nmor1, nmor2} (although some estimates were given in Ref.~\cite{zoran}). Since the effects are employed in optically-pumped magnetometers, this may have important consequences, leading to systematic errors of the devices in fundamental searches or spectroscopic measurements.

In this Paper, we present both experimental and theoretical (numerical) studies of BSS in a Rb vapor system subjected to static and oscillating/rotating magnetic fields. The effect is studied by means of radio-frequency nonlinear magneto--optical rotation (RF NMOR) \cite{ledbetter, zigdon}. In this kind of experiment, RWA is typically applied with respect to the oscillating magnetic field and BSS disregarded. Important features of our system are: a possibility of application of either rotating or oscillating magnetic field under otherwise the same experimental conditions, generation of narrow ($\approx$1~Hz) RF NMOR resonances, and measurements at ultra-low frequencies ($\omega/2\pi = 1$~Hz to $100$~Hz). The first aspect is essential for elimination of systematic errors, while the second and third are relevant because of inverse proportionality of the shift to the magnitude of static (DC) magnetic field [Eq. \eqref{bs}]. Precision of our method, enhanced by the small value of $\Omega_0$, enables utilization of much weaker alternating (AC) fields than in previous experiments. Hence, in our case the perturbative approach derived in Ref.~\cite{cohen,shirley,hannaford} is valid. Exerting strong RF fields to reach the limit where perturbation method is no longer credible (i.e. $\Omega_{RF} \gtrsim 2.4  \Omega_0$) leads to broadening and deformation of resonance lineshapes, and thus reduced precision of the measurement \cite{cohen,hannaford}. Moreover, we discuss another effect, leading to deformation of the observed resonances, which stems from the complexity of the system ($F\geq 1$). 

\section{Numerical approach}
We investigate BSS of NMOR resonances using the density-matrix formalism and a model atomic system ($F = 1$ ground state and $F' = 0$ excited state) interacting with linearly polarized light and immersed in AC and DC magnetic fields. This is the simplest system enabling discussion of the NMOR signals. The Hamiltonian of the system is presented in a rotating frame corresponding to the optical frequency $\Omega$ and we apply RWA in the optical-field domain. In the $|F,m_F \rangle$ base, with a quantization axis along the light propagation direction and perpendicular to the static magnetic field, the Hamiltonian is given by
\begin{equation}
{H} = {H}_0+{H}_E+{H}_B,
\label{ham}
\end{equation}
\begin{equation}
 H_0 = \hbar \Delta \begin{bmatrix} 0   & 0 &0 &0 \\
    0     & 0 & 0 & 0\\
   0  & 0 & 0  &0 \\
    0       & 0 & 0 & 1
    \end{bmatrix},   
  \quad H_E= \frac{i \hbar \beta }{2}
\begin{bmatrix}
  0 & 0 &0 &-1 \\
     0 & 0 & 0 & 0\\
   0 & 0 & 0 &-1 \\
    1       & 0 & 1 & 0 
\end{bmatrix},
\end{equation}
where $ \Delta = \omega_0-\Omega$ is the laser frequency detuning  and $\beta$ is the optical Rabi frequency. The form of $H_B$ depends on the AC field. For  $\boldsymbol{B_{lin}}=\hbar/ (g \mu_B)[0, \Omega_0, \Omega_{RF} \cos(\omega t)]^T$, $H_B$ equals to
\begin{equation}
H^{lin}_{B}= 
\hbar \begin{bmatrix}
 \Omega_{RF}\cos(\omega t)  & \frac{-i\Omega_0}{\sqrt{2}} & 0 &0 \\
    \frac{i\Omega_0}{\sqrt{2}}     & 0 & \frac{-i\Omega_0}{\sqrt{2}} & 0 \\
   0  & \frac{i\Omega_0}{\sqrt{2}} &  -\Omega_{RF}\cos(\omega t) & 0 \\
    0  & 0 & 0 & 0 
\end{bmatrix}.
\label{linh}
\end{equation}
In the case of rotating magnetic field $\boldsymbol{B_{rot}}=\hbar/ (g\mu_B)[ \frac{\Omega_{RF}}{2} \sin(\omega t), \Omega_0, \frac{\Omega_{RF}}{2} \cos(\omega t)]^T$, $H_B$ is given by
%\par\noindent\rule{\dimexpr(0.5\textwidth-0.5\columnsep-0.4pt)}{0.4pt}%
%\rule{0.4pt}{6pt}
\begin{widetext}
\begin{equation}
H^{rot}_{B}= 
\hbar \begin{bmatrix}
\frac{ \Omega_{RF}}{2}\cos(\omega t)  & \frac{-i\Omega_0}{\sqrt{2}}+ \frac{ \Omega_{RF}}{2\sqrt{2}}\sin(\omega t) &0 &0 \\
     \frac{i\Omega_0}{\sqrt{2}}+ \frac{ \Omega_{RF}}{2\sqrt{2}}\sin(\omega t)     & 0 &  \frac{-i\Omega_0}{\sqrt{2}}+ \frac{ \Omega_{RF}}{2\sqrt{2}}\sin(\omega t) & 0 \\
   0  & \frac{i\Omega_0}{\sqrt{2}}+\frac{ \Omega_{RF}}{2\sqrt{2}}\sin(\omega t) &  -\frac{\Omega_{RF}}{2}\cos(\omega t) & 0  \\
    0  & 0 & 0 & 0 
\end{bmatrix}.
\label{roth}
\end{equation}
\end{widetext}
%\vspace{\belowdisplayskip}\hfill\rule[-6pt]{0.4pt}{6.4pt}%
%\rule{\dimexpr(0.5\textwidth-0.5\columnsep-1pt)}{0.4pt}
The density-matrix evolution is described by the Liouville equation
 \begin{equation}
 i \hbar \frac{d \rho}{dt}= [H,\rho] - i\hbar \lbrace\Gamma, \rho\rbrace + i \hbar \Lambda,
 \label{liou}
 \end{equation}
where $\Gamma$ and $\Lambda$ are the relaxation and repopulation matrices:
 \begin{equation}
 \Gamma=
\begin{bmatrix}
\gamma&0&0&0\\
0&\gamma&0&0\\
0&0&\gamma&0\\
0&0&0&\gamma+\gamma_e
\end{bmatrix},
 \; \Lambda=\frac{\gamma + \gamma_e \rho_{00}}{3}\begin{bmatrix}
1&0&0&0\\
0&1&0&0\\
0&0&1&0\\
0&0&0&0
\end{bmatrix},
 \end{equation}
 where $\rho_{00}$ is the excited state population.
No atomic kinematics, like thermal motion, is taken into account.
Equation \eqref{liou} describing time evolution of the density matrix is solved numerically \cite{pakiet} and rotation angle $\phi$ of linearly polarized light acquired along the optical path is calculated as
 \begin{equation}
 \phi \propto \mathrm{Im}(\rho_{-10}-\rho_{10}),
 \end{equation}
where $\rho_{-10}$ ($\rho_{10}$) is the non-zero optical coherence between the $|1,-1\rangle$ ($|1,1\rangle$) ground state and the $|0,0\rangle$ excited state. %\footnote{%Note the difference from typical Faraday geometry where magnetic field is parallel to light propagation direction and $ \phi \propto \mathrm{Re} (\rho_{|1,-1\rangle|0,0\rangle}+\rho_{|1,1\rangle|0,0\rangle})$.}.
Oscillation or rotation of $\boldsymbol{B_{RF}}(t)$ results in modulation of rotation angle $\phi$. To reproduce measurements of the modulated observables with the phase sensitive detection, solutions of differential equations are integrated over the period $T={2 \pi}/{\omega}$ after the time $t_S$ necessary for the system to approach a steady state
 \begin{equation}
\phi_{p} (\omega)+ i \phi_{q}(\omega) =\int_{t_{S}}^{t_{S}+T} \phi(\tau) \exp( i \omega \tau) d\tau.
\end{equation}
Subscripts \textit{p} and \textit{q} correspond to the signal measured in phase and in quadrature with the AC magnetic field. 

\begin{figure}
\centering
\includegraphics[width=\linewidth]{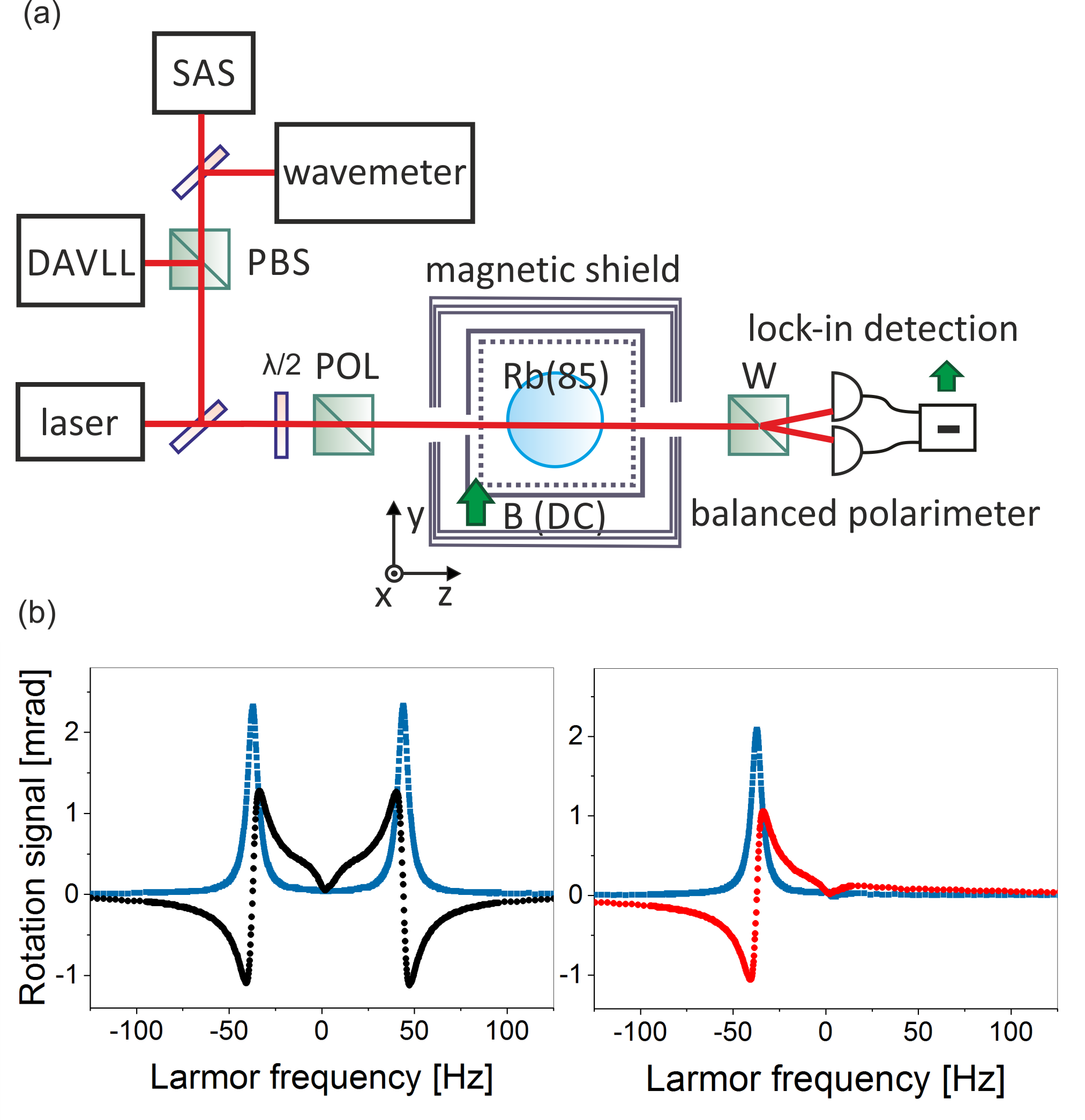} 
\caption{Setup schematics and RF NMOR signals at low $\Omega_{RF}$. (a) $^{85}\mathrm{Rb}$ ensemble is pumped and probed by linearly polarized light tuned to the low-frequency wing of the Doppler-broadened $F = 3 \rightarrow F'$ transition. Coil producing magnetic field in the $y$ direction is fed with a constant current, while the $x$ and/or $z$ coils are fed with both direct and alternating current for compensation and generation of rotating/oscillating magnetic field. The polarization rotation is measured with a balanced polarimeter and a lock-in amplifier. DAVLL stands for the dichroic atomic vapor laser lock, SAS is the saturated absorption spectroscopy, PBS is the polarizing beam splitter, W is the Wollaston prism, ${\lambda}/{2}$ denotes the half-wave plate used with a Glan--Thompson polarizer (POL) for the light power adjustment. The rotation signals are recorded with a computer (not shown). (b) The in-phase (black/red or grey) and quadrature (blue or dark grey) components of the recorded resonances for oscillating (left) and rotating (right) magnetic fields. The AC field frequency is held constant while the static magnetic field is scanned.}
\label{setup}
\end{figure}

\section{Experimental setup}
The central part of our experimental setup [Fig. \ref{setup}(a)] is a spherical, paraffin-coated vapor cell (of $10$ cm diameter) filled with isotopically enriched $^{85} \mathrm{Rb}$ (relaxation time $T_1 \approx 1$~s).  The cell is surrounded by a three-layer magnetic shielding made of $\mu$-metal and a ferrite cube as the innermost layer. A set of magnetic-field coils generating homogeneous magnetic fields in all three directions is mounted inside the shield. The leading field is set along the $y$ axis. Other coils serve for passive compensation of residual fields and application of a variable magnetic field. In case of oscillating field, one of the perpendicular coils ($x$ or $z$) is fed with a sinusoidal signal. The rotating field is produced when $\cos(\omega t)$ and $\sin(\omega t)$ signals are applied to $x$ and $z$ coils, respectively.

Linearly polarized laser beam tuned to the D1 ($795$~nm) line of rubidium creates spin alignment and monitors atomic polarization changes due to propagation through a coherently prepared atomic media. It is tuned to a wing of the Doppler broadened $F = 3 \rightarrow F'$ transition (detuned -260~MHz from the $F=3\rightarrow F'=2$ transition). The laser light is stabilized with dichroic atomic vapor laser lock (DAVLL) \cite{davll1, davll2} and saturated absorption spectroscopy (SAS) serves as frequency reference. Typical light power prior the cell is 20~$\mu$W (in paraffin-coated cells, the power is averaged over the volume \cite{zhivun}). Magneto--optical (polarization) rotation of light is detected by a balanced polarimeter. The signal is demodulated with a lock-in amplifier operating at the RF field frequency typically between $1$ and $100$ Hz. 

The RF NMOR signals obtained with sweeping of the static magnetic field and fixed RF-field frequency are presented in Fig.~\ref{setup}(b). The Larmor frequency corresponding to the static field $\Omega_0$ is extrapolated from an independent magnetic-field calibration (high frequency regime, low AC-field amplitude). The oscillating field generates two resonances at $\pm\Omega_r$ and only a single resonance at $\Omega_r$ occurs with the rotating field. The in-phase component of the observed resonances is characterized by the dispersive shape, while its quadrature counterpart is absorptive. Width of both resonances is determined by the transverse ground-state relaxation rate. A small dip around $\Omega_0 = 0$ is considered as an experimental singularity caused by $1/f$ noise in the system \footnote{Modeling without the noise does yield signals without central zero-field feature, but adding pink noise reproduces faithfully the signals of the Fig. 1(a). it was also experimentally verified that the width of the central dip increases with the vapor temperature (which confirms thermal orign of the noise).}.

The drawback of the scheme, where RF field frequency is fixed, is is the necessity of scanning the magnetic-field value. Due to the change in operation conditions of the current source, magnetic-field scanning may introduce uncertainties to the measurements (magnetic-field drifts). To avoid this problem, further results are taken with a constant magnetic field (more stable operation conditions) and the RF-field frequency $\omega$ swept around the resonance frequency $\omega_r$. However, a restriction of such a scheme is the ability of scanning only positive frequencies of the RF field.

\begin{figure}
\centering
\includegraphics[width=\linewidth]{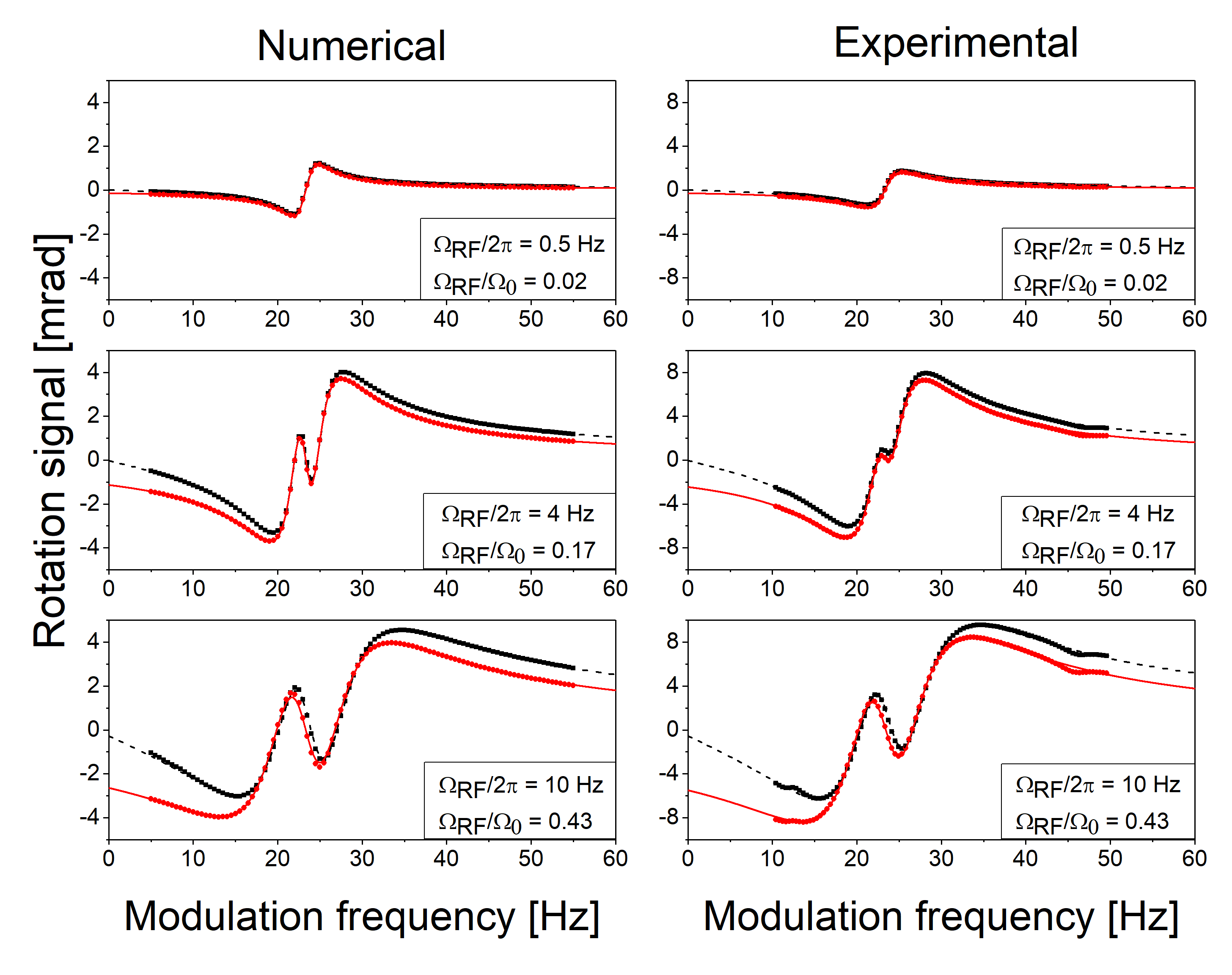} 
 \caption{Numerical (left) and measured (right) NMOR signals for oscillating (black squares) and rotating magnetic fields (red or grey dots) along with fitted Majorana reversal NMOR functions (dashed and solid lines respectively)[Eq. (11)] for three different AC magnetic field strengths indicated in right bottom corner of each graph. Static magnetic field was set to $\Omega_0/2\pi = 23$ Hz. Residual magnetic field is responsible for appearance of weak resonances at $\omega_{r}/2$ and $2 \omega_{r}$ observed in experimental data for high $\Omega_{RF}$. Statistical error bars of experimental data are smaller than the dot/square size. Notice the difference in vertical scale for experiment and numerical solutions. It is due to simplicity of simulated model.}
\label{linexp}
\end{figure}
\section{Results}
Figure \ref{linexp} presents numerical and experimental signals (in-phase component) generated with the oscillating and rotating fields at three AC-field amplitudes (expressed in terms of Rabi frequency $\Omega_{RF}$ \footnote{$\Omega_{RF}$ was calibrated with separate measurement. $V(t) = V_{RF}+V_{RF} \cos(\omega t)$ signal was applied to the $z$ coil and resonance frequency $\Omega_{RF}/ \pi$ measured.}). At low $\Omega_{RF}$, the shape of the resonance signal is purely Lorentzian. When the AC field becomes stronger the resonance broadens and a new narrow structure develops in its center. This effect, characteristic for systems with $F\geq 1$, was observed and described first in Ref. \cite{brossel} and its rigorous calculation was derived by Majorana \cite{majorana}. It stems from quantum interference of one- and three-photon processes, each contributing to the resonance with a different linewidth and amplitude depending on $\Omega_{RF}$ \cite{CTT}. The resonance lineshapes, including the Majorana reversal for RF NMOR, were calculated in Ref. \cite{wais} and we use them in our fitting procedure in the following form

\begin{equation*}
\phi_p (\omega)=  \frac{\phi_0 \omega \Omega_{RF}[\gamma^2-2\Omega^2_{RF}+4(\omega-\omega_r)^2]}{[\gamma^2+\Omega^2_{RF}+(\omega- \omega_r)^2][\gamma^2+4\Omega^2_{rf}+4(\omega- \omega_r)^2]},
\end{equation*}

\begin{equation}
\phi_q (\omega)= \frac{\phi_0 \Omega_{RF}[\gamma^2+\Omega^2_{RF}+4(\omega-\omega_r)^2]}{[\gamma^2+\Omega^2_{RF}+(\omega- \omega_r)^2][\gamma^2+4\Omega^2_{RF}+4(\omega- \omega_r)^2]},
\label{eq:fitting}
\end{equation}
where $\phi_0$ is the amplitude of the magneto-optical rotation. 

The correct description of RF NMOR with oscillating magnetic field requires taking into account the existence of the second resonance at negative frequencies (not observed directly in the experiment). Omission of the resonance would lead to systematic algebraic shifts.

%In case of the oscillating field our fitting function consists of resonances at both frequencies $\pm \omega_r$. This way an algebraic resonance attraction/repulsion due to the existence of implicit resonance at the negative frequency is eliminated. Nevertheless, trying to impose a single resonance fit to oscillating-field data at low $\Omega_0$ would result in indisputable fit model failure.

Figure~\ref{linexp} shows that for stronger AC fields, small features at $\omega_r/2$ and $2\omega_r$ are observed in experimental signals. We attribute the features to the residual magnetic fields and fit them with small absorptive resonances with predefined positions ($\omega_r/2$ and $2\omega_r$) and a given width. 

Since the residual fields are also responsible for small mixing betweem two lock-in outputs, we fit the experimental data with a combination of the resonance shapes given by Eqs.~\eqref{eq:fitting}. In case of the rotating-field-induced resonance, residual fields are exclusively responsible for the mixing. When the oscillating-field-induced resonance is considered, an additional phase mixing is produced by counter-rotating component of the AC field (interference of the measured resonance with the tail of the resonance at $-\omega_r$).

For a quantitative characterization of the Bloch--Siegert effect we foused on the difference between resonance frequencies when the driving field is either oscillating or rotating, $\Delta \omega = \omega^{rot}_r - \omega^{lin}_r$, and regarded it as the measure of BSS. In Fig. 3, the results of $\Delta \omega$ versus $\Omega_{RF}$ are compared with numerical calculations at three different DC field strengths. Maximum value of the applied RF field, $\Omega_{RF}= 12$ Hz $< 0.5 \Omega_0/2\pi$, allows us to limit the analysis to the lowest-order term of the perturbative expression \cite{shirley, hannaford}.
 
 \begin{figure}[t!]
\centering
  \includegraphics[width=\linewidth]{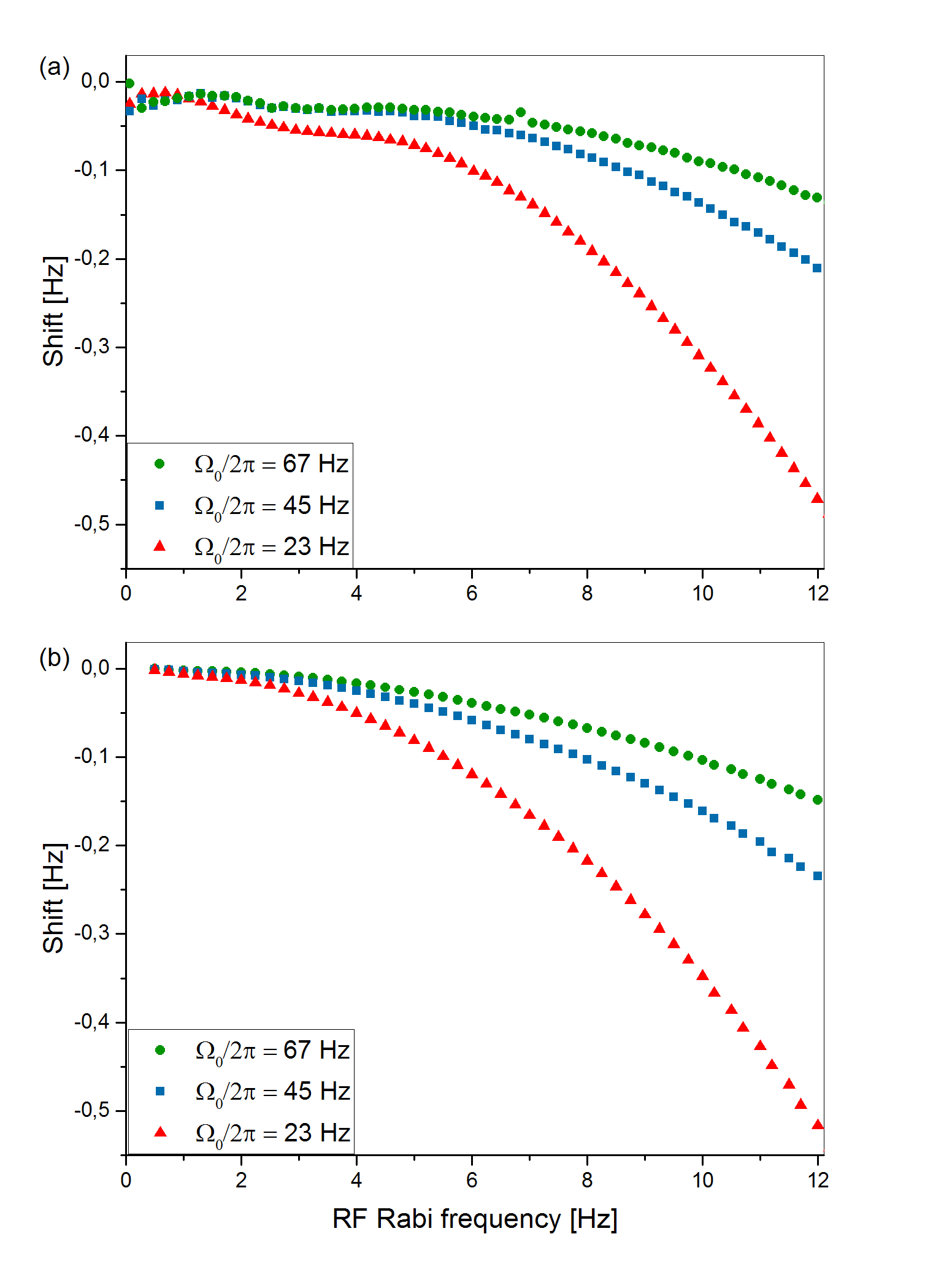}
\label{fit}  
\caption{Results of (a) experiment and (b) numerical calculations performed for three different Larmor frequencies. The shift depends quadratically (first order approximation) on RF Rabi frequency and scales inversely with the static magnetic field. Note the negative value of the shift, according to the convention of  Eq. \eqref{bs}. The minus sign reflects conversion from the static magnetic field sweeping to the frequency sweeping of AC field.}
\end{figure} 
 
As shown in Eq. \eqref{bs}, both the measured and calculated shift values appear inversely proportional to the intensity of the static magnetic field and increase quadratically with the amplitude of the modulated magnetic field. This is confirmed with our results. Some anomaly is seen around $\Omega_{RF}/2\pi =2.5$ Hz, where experimental BSS value seem to be amplified, and needs yet to be investigated.
 
In Fig.~4, the resonance position ($\Omega_0/2\pi=45$~Hz) as the function of $\Omega_{RF}$ is shown for both rotating and oscillating AC field. Solid red (or grey) and black dashed lines represent numerical solutions while experimental datapoints are presented as dots and squares. As shown, BSS measured experimentally is slightly overestimated by the numerical results. The small systematic shift of the rotating-field results is believed to be caused by the magnetic-coil asymmetry (non-zero oscillating component along the DC field) and hence smaller RF-field modulating atomic polarization. Additionally, inequality of two counter-rotating magnetic-field components in the rotating-field measurements, may lead to non-zero signal at $-\omega_r$ and hence algebraic shift of the resonance.

 \begin{figure}[t!]
\centering
  \includegraphics[width=\linewidth]{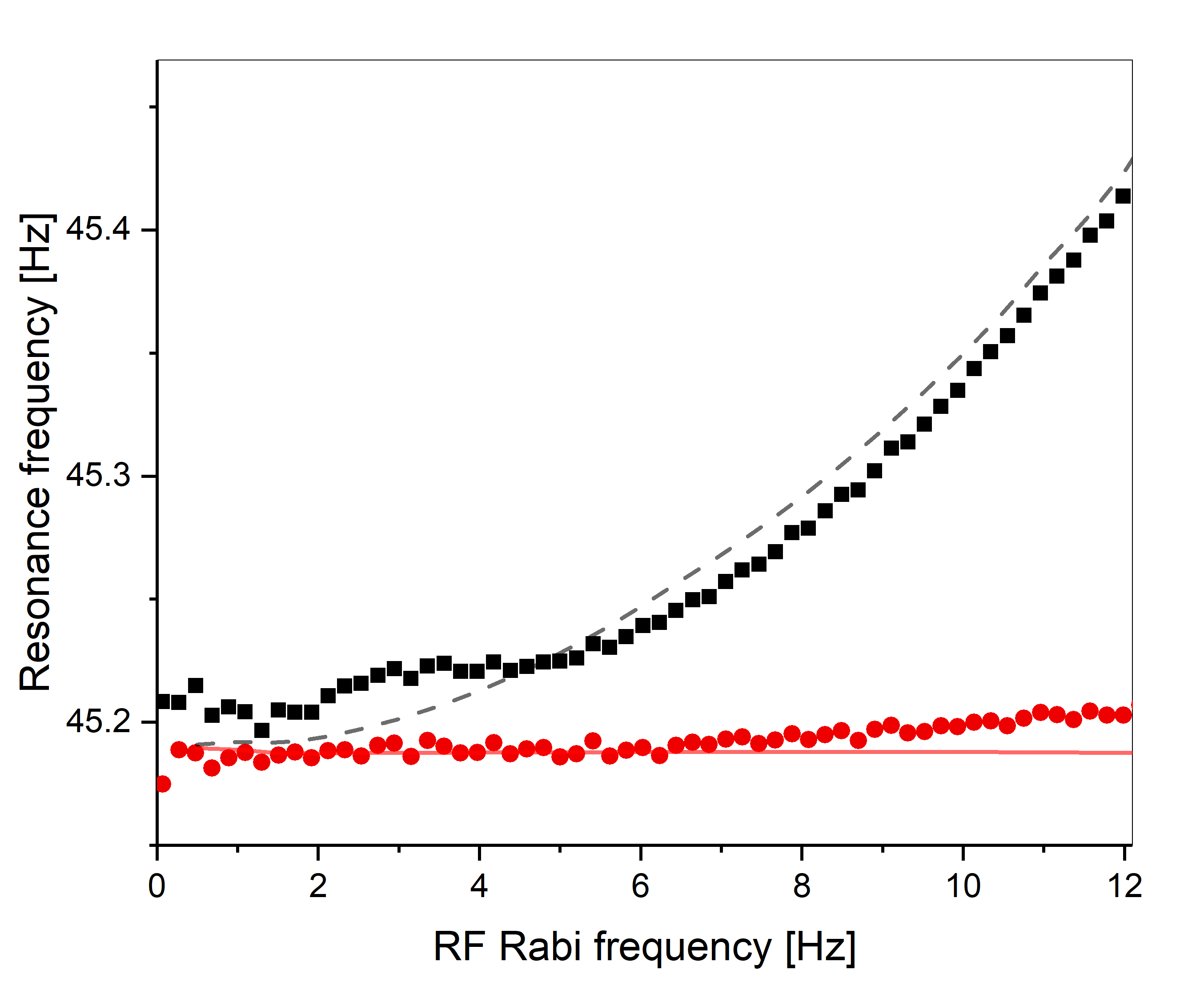}
\label{fit}  
\caption{Resonance position in function of $\Omega_{RF}$ for oscillating- (black) and rotating-field induced (red or grey) resonances at static magnetic field $\Omega_0/2\pi = 45$ Hz. Solid and dashed lines represent numerical solutions while squares and dots are experimental datapoints.}
\end{figure}

Despite that, experimental datapoints do not deviate strongly from theoretical curves. Particularly, the resonance position barely depends on the AC field Rabi frequency for rotating-field-induced resonances, which illustrates elimination of BSS.

\FloatBarrier
\section{Conclusions}
In conclusion, by performing precision magneto--optical rotation measurements we have characterized BSS in a multilevel system of Rb atoms in a wide range of parameters. The most prominent effects is two-component resonance structure at $-\omega_r$ and $+\omega_r$ for the oscillating field and the single component at $+\omega_r$ for the rotating one.
BSS was studied as a function of such parameters as static and RF-field strength. We carefully investigated various features of the observed resonances, e.g., the Majorana reversal, phase mixing, and extra resonances due to field inhomogenity. All the features were confirmed with numerical simulations based on the density-matrix calculations.
%\footnote{The misalignment in numerical results may be provided by proper modification of $H_B$ in Eq. \eqref{linh} and Eq. \eqref{roth}.}. 

With our investigations on BSS, we demonstrated limitations of the rotating-wave approximation in magneto-optical effects. We showed that the resonance position may depend not only on such parameters of AC magnetic field as frequency and amplitude, but also on the position at which the resonance is observed (the Larmor frequency). Our measurements were performed in the regime of weak magnetic fields, where the shift depends on the quadratically on the magnetic-field amplitude. Manifestations of multi-photon processes were also observed. These results improve our understanding of the processes associated with light-atom interaction in the presence of the magnetic field. It should be stressed that the shift becomes an important problem in optically-pumped magnetometers. In such a case, the existence of the counter-rotating component of the magnetic field may lead to systematic errors in the measurements of the magnetic fields. This may be crucial in experiments where measurements of good accuracy are required (permanent electric dipole moment experiments \cite{nEDM}, searches for time-variation of fundamental constants \cite{alfa}, NMR \cite{trahms}, etc.). As shown in this paper, BSS can be avoided by application of rotating, resonant magnetic fields.

\vspace{10pt}
\begin{acknowledgments}
The authors would like to acknowledge stimulating discussion with D. Budker and D. F. Jackson Kimball. S.P. and W.G. would like to recognize fundings from the Polish National Centre of Science (grants 2015/19/B/ST2/02129, 2012/07/B/ST2/00251, 2016/21/B/ST7/01430).

This article is dedicated to the memory of prof. J. Zachorowski.
\end{acknowledgments}

% Create the reference section using BibTeX:
%

%\end{thebibliography}

\end{document}